\documentclass[epsf,useAMS,usenatbib]{mn2e}

\usepackage{lscape}
\usepackage{graphicx}
\usepackage{natbib}
\usepackage{graphics}

\def \vhel{\ifmmode{~V_{{\rm HEL}}}\else{~$V_{{\rm HEL}}$}\fi}
\def \vsys{\ifmmode{~V_{{\rm SYS}}}\else{~$V_{{\rm SYS}}$}\fi}
\def \HA {\ifmmode{{\rm\H}\alpha}\else{${\rm\ H}\alpha$}\fi}
 
\def \msun{\ifmmode{{\rm\ M}_\odot}\else{${\rm\ M}_\odot$}\fi}
\def \myr{\ifmmode{{\rm\ M}_\odot{\rm\ yr}^{-1}}
         \else{${\rm\ M}_\odot$ yr$^{-1}$}\fi}
\def \mdot{\ifmmode{\dot{M}}\else{$\dot{M}$}\fi}
\def \tena#1 #2 {\ifmmode{#1 \times 10^{#2}}\else{$#1 \times 10^{#2}$}\fi}
\def \kms{\ifmmode{~{\rm km\,s}^{-1}}\else{~km s$^{-1}$}\fi}

\title[Cosmic ray confinement in fossil cluster bubbles]{
Cosmic ray confinement in fossil cluster bubbles}
\author[M. Ruszkowski et al.]{M. Ruszkowski,$^1$\thanks{E-mail:
mr@mpa-garching.mpg.de (MR)}  
T.A. En{\ss}lin$^1$, M. Br{\"u}ggen$^2$, M.C. Begelman$^{3,4}$
\& E. Churazov$^{1,5}$\\
$^1$Max Planck Institute for Astrophysics, Karl-Schwarzschild-Str. 1,
85741 Garching, Germany\\
$^2$Jacobs University Bremen, Campus Ring 1, Bremen, Germany\\
$^3$JILA, University of Colorado at Boulder, CO 80309-0440, USA\\
$^4$Department of Astrophysical and Planetary Sciences, University of
Colorado, Boulder, CO 80309-0391, USA\\
$^5$Space Research Institute, Profsoyuznaya str. 84/32, Moscow 117997, Russia\\
}
\begin{document}

\date{Submitted 2007 May}

\pagerange{\pageref{firstpage}--\pageref{lastpage}} \pubyear{2006}
\maketitle
\label{firstpage}
\begin{abstract}
Most cool core clusters of galaxies possess active galactic nuclei (AGN) in their centers. These AGN
inflate buoyant bubbles containing  non-thermal radio emitting particles. If such bubbles efficiently confine cosmic rays (CR) then
this could explain ``radio ghosts'' seen far from cluster centers.  We simulate the diffusion of cosmic rays from buoyant bubbles inflated by AGN. Our simulations include the effects of the anisotropic particle diffusion introduced by magnetic fields.  Our models are consistent with the X-ray morphology of AGN bubbles, with disruption being suppressed by the magnetic
draping effect. We conclude that for such magnetic field topologies, a substantial fraction of cosmic rays can be confined inside the bubbles on buoyant rise timescales even when the parallel diffusivity coefficient is very large. For isotropic diffusion at a comparable level, cosmic rays would leak out of the bubbles too rapidly to be consistent with radio observations. Thus, the long confinement times associated with the magnetic suppression of CR diffusion can explain the presence of radio ghosts. We show that the partial escape of cosmic rays is mostly confined to the wake of the rising bubbles, and speculate that this effect could: (1) account
for the  excitation of the H$\alpha$ filaments trailing behind the bubbles in the Perseus cluster, (2) inject entropy into the metal enriched material being lifted by the bubbles and, thus, help to displace it permanently from the cluster center and (3) produce observable $\gamma$-rays via the interaction of the diffusing
cosmic rays with the thermal intracluster medium  (ICM).

\end{abstract}


\section{Introduction}
The discovery of X-ray cavities in the ICM that coincide with radio emission (e.g.,
Blanton et al. 2001, Boehringer et al. 2003), sound waves and weak shocks (Fabian et al. 2003a) provided clear observational support for the idea that AGN are the main heating source that could offset radiative cooling and prevent ``cooling catastrophes'' in clusters of galaxies.
There is growing consensus that overall heating and cooling balance may indeed be achieved. This is based on the analysis of 
statistically significant samples of ICM bubbles (Birzan et al. 2006) as well as other arguments. However, the actual mechanism 
by means of which the mechanical energy of AGN is delivered to the thermal ICM remains elusive.\\
\indent
The bubbles that supply energy to the ICM are known to be filled with  magnetized, relativistic, radio-emitting, non-thermal plasma and, thus, the diffusion of cosmic rays (CR) from magnetic bubbles and their interaction with the ICM is one of the key physical mechanisms that needs to be explored in the context of AGN feedback in clusters. These processes are of high astrophysical
relevance to a number of long-standing problems. If the diffusion rates are relatively high, cosmic ray diffusion may
explain the absence of strong "cooling flows", the origin of cluster radio halos and, as  we suggest in this paper, 
the origin of the excitation of H$\alpha$ filaments in clusters 
\footnote{A similar possibility was independenty suggested by Sanders \& Fabian (2007)} and 
the metal distribution in clusters. If, on the other hand,  CR remain largely confined to the bubbles as they rise buoyantly, they may explain the presence of radio relic sources. Such electrons will be protected against Coulomb losses and are thus able to produce a significant non-thermal Comptonisation signature of the CMB.\\
\indent
The effects of CR diffusion in clusters have been previously studied by, e.g., Mathews \& Brighenti (2007), who used a one-dimansional
 model  to put upper limits on the effective diffusivity rate of cosmic rays from AGN-inflated bubbles. These effects were also addressed by  Jones et al. (1999) and Tregillis et al. (2004),  who performed 3D magnetohydrodynamical (MHD) simulations focusing on a different regime (high Mach number jets). Cosmic ray diffusion was also incorporated recently into cosmological structure formation simulations by Miniati et al. (2001) (see also Miniati 2001 for the description of the COSMOCR code) and by En{\ss}lin et al. (2007) and by Jubelgas et al. (2006) using the GADGET code.\\
\indent
In this paper we consider three-dimensional MHD simulations of AGN-inflated bubbles in their buoyant stage. We go beyond previous treatments of this feedback mechanism and include the anisotropy of cosmic ray diffusion introduced by magnetic fields in the bubbles.
In the next section we describe the methods employed to simulate anisotropic diffusion. In section 3 we present our
results and in the last section we discuss the astrophysical consequences  of our findings.

\section{Simulation details}

\subsection{The code}
The simulations were performed with the {\it PENCIL} code (Dobler et al. 2003, Haugen et al. 2003, Brandenburg et al. 2004, Haugen et al. 2004). {\it PENCIL} is a highly accurate grid code that is sixth order in space and third order in time. It is particularly suited for compressible turbulent MHD flows. Magnetic fields are implemented in terms of a vector potential so the field remains solenoidal throughout the simulation. The code is memory-efficient, uses Message-Passing Interface and is highly parallel. 

\subsection{Initial conditions}

The details of the numerical setup pertaining to the thermodynamical state of the cluster atmosphere and magnetic
fields were described in detail in Ruszkowski et al. (2007) (hereafter R07). Here we summarize only the essential points and differences. The distribution of magnetic fields in all runs presented here corresponds to the ``draping case'' whereby the characteristic lengthscale of magnetic field fluctuations exceeds the bubble size. The magnetic pressure is much smaller than the gas pressure (typical plasma $\beta$ parameter is of order 20; see R07 for the discussion of this choice). Even though such a magnetic field is formally ``dynamically unimportant'', it
nevertheless strongly affects the dynamical state of the bubbles by preventing their disruption via the Rayleigh-Taylor and Kelvin-Helmholz instabilities, thus making their morphologies consistent with X-ray observations.  Bubbles were initially in total (i.e., CR and thermal gas) pressure equilibrium with the ambient ICM. For numerical reasons we chose the CR pressure to be such that its volume-averaged value was comparable to the thermal pressure. More specifically, the cosmic ray pressure fraction $f_{\rm CR}$ was distributed inside the bubble according to $f_{\rm CR}(r)=(1-f_{\rm gas}\cos(0.5\pi r/r_{b})$, where $f_{\rm gas}=0.1$
is the pressure fraction due to the thermal gas at the centre of the bubble, $r_{b}=12.5$ kpc is the bubble radius and $r$ is the distance from the bubble centre. We assumed that the cosmic ray contribution to pressure outside the bubble is negligible. The bubbles were made buoyantly unstable. At each point within the bubble volume, the gas density was $\rho_{\rm gas}=\rho_{\rm ICM}(1-f_{\rm CR})/x$ and the gas temperature $T=T_{\rm ICM}x$, where $x=10$. The temperature of the ambient ICM was constant and equal to 10keV (note that high values of temperature lower the effective viscosity of the code as compared to the Braginskii value). A constant value for the ambient temperature also makes it easlier to identify the volume occupied by the bubble (i.e., there is no need to introduce a  dynamically passive fluid to define the bubble boundaries; we use this fact in the analysis of our simulations below).\\
\indent
The resolution of our simulations was 200$^3$. We used the following code units in this work: code time unit was $3.3\times 10^6$ yr, code
length unit was 1 kpc and density unit $10^{-24}$ g cm$^{-3}$. In these units, the diffusivity $K=32.4(K_{0}/3\times 10^{30}{\rm
cm}^{2}{\rm s}^{-1})$ code units, where $K_{0}$ denotes the diffusivity in cgs units. 

\subsection{Cosmic ray diffusion}
Cosmic rays are incorporated into the code via a two-fluid model. The gas and cosmic ray plasmas have adiabatic indices equal to $\gamma=5/3$ and $\gamma_{\rm cr}=4/3$, respectively. Cosmic rays are coupled to the gas and act on it via their pressure gradient.\\
\indent
Diffusion of relativistic particles in {\it PENCIL} takes into account the effects of anisotropy introduced by magnetic fields
(diffusion in the direction perpendicular to the field lines is slower than along the field lines). \\
\indent
The standard diffusion equation violates causality by allowing for infinite particle propagation speeds. The diffusion method implemented in the {\it PENCIL} code (Snodin et al. 2006) improves on the standard Fick diffusion law and takes into account the requirement that particles should propagate at finite  speeds.  This is achieved by expanding the Boltzmann equation in the relevant smallness parameter and retaining terms up to second order (see Gambosi et al. 1993 for the discussion of charged particle transport). This treatment is an extension of the method of Hanasz \& Lesch (2003), who implemented CR diffusion in the ZEUS code, that included anisotropic diffusion but did not consider departures from the standard diffusion law. The relevant smallness parameter in the expansion of the Boltzman equation is the Strouhal number $St\equiv (K_{\parallel}\tau)^{1/2}/l$, where $(K_{\parallel}$ 
is the diffusion coefficient along the magnetic field lines, $l$ is the characteristic lengthscale of the initial magnetic structure and $\tau$ is the damping time of the non-Fickian contribution to diffusion; Landau \& Lifshitz 1987). This approach leads to the generalization of the diffusion equation to  the ``telegraph'' equation that reduces to the standard diffusion equation in the limit of $St\rightarrow 0$ but becomes a wave equation
for $St\rightarrow\infty$. A non-Fickian diffusion model has also been  derived in the context of turbulent diffusion of passive scalars  by Blackman \& Field (2003) and  confirmed in numerical experiments (Brandenburg, K{\"a}pyl{\"a} \& Mohammed 2004).\\
\indent
A very rough estimate of the diffusion coefficient can be obtained from a dimensional argument based on the radius of the inner X-ray cavities in the Perseus cluster $r\sim 5r_{5}$ kpc and the time it takes to inflate the bubble $t\sim 10^7t_{7}$ yr  Mathews \&
Brighenti 2007). These values yield $K\la 7.5\times 10^{29}r^{2}_{5}/t_{7}$ but the estimate is somewhat uncertain. This value is consistent with the theoretical estimates of En{\ss}lin (2003) $\kappa_{\parallel}\sim 2\times 10^{29}
E_{10}^{1/3}r_{5}^{2/3}B_{1}^{-1/3}\delta^{-1}$cm$^2$s$^{-1}$, where $E_{10}$ is the energy of radio emitting electrons in units of 10 GeV,
$B_{1}$ is the magnetic field in $\mu$Gauss, and $\delta$ is a parameter of order unity or less that depends on the power spectrum of magnetic field fluctuations and the ratio of particle scattering frequency to gyro-frequency. Assuming that $\tau\sim r/v_{A}$,
where $v_{A}$ is the Alfv\'en speed, $\tau$ is of order unity in code units for the typical parameters in our simulations. As $\tau$ is essentially a free parameter of this model we adopt this value here. Guided by the above estimates for the diffusion coefficient and the damping time, we estimate the Strouhal number to be $St\sim 0.5 (\tau K_{\parallel}/K)^{1/2}(r/r_{b})^{-1}$, where
$r_{b}=12.5$ is the bubble radius in code units and $K=32.4$ is the reference diffusivity in code units  (in cgs units it is $3\times 10^{30}$cm$^{2}$s$^{-1}$).  The code requires some magnetic diffusivity, $\eta$, viscosity, $\nu$, and an isotropic contribution to cosmic ray diffusivity, $K_{\rm iso}$, to ensure numerical numerical
stability. These coefficients can be lowered at the expense of increasing numerical resolution. Except where explicitly stated, in our simulations we adopted $\eta=\nu=K_{\rm iso}=0.07$. As we intend to study the effect of the anisotropy of cosmic ray diffusion, we want to keep the ratio of $K_{\parallel}/K_{\rm iso}$ as large as possible. We also want to be conservative in our assessment of the importance of the diffusion and, therefore, consider high values of diffusivity. We stress that for high diffusivities, simulations would not be feasible if not for the fact that the code uses a  non-Fickian approach, i.e., effectively limits the CR propagation speed and prevents the timestep from decreasing to unacceptably low values.

\section{Results}
Figure 1 shows snapshots from our simulations in the high diffusivity regime.
The first column shows the natural logarithm of thermal gas density in the cluster, while the second and third columns show the logarithm of the CR energy density for two different assumptions about the isotropy of diffusion. The second column corresponds to the case where diffusion in the direction perpendicular to the local magnetic field is suppressed (i.e., $K_{\perp}=0$, $K_{\parallel}=K$), while the last column is for the case where the diffusion coefficients of
cosmic rays in the direction parallel and perpendicular to the local direction of magnetic field lines are identical. As the density distribution looks quite similar in both the isotopic and anisotropic cases, we show only the density map for the anisotropic diffusion case. The reason for this similarity is that the pressure support in cosmic rays is comparable to that in the thermal component (for numerical stability reasons) and, thus, even when all cosmic rays
diffuse out of the bubble, its radius diminishes only by a modest
amount of $\sim 1-0.64^{-1/(3\gamma_{\rm cr})}\sim 10\%$.\\
\indent
The first column in Figure 1 demonstrates that the ``draping effect'' of the magnetic field prevents the disruption of the bubble as it rises
buoyantly in the cluster atmosphere. We also note that the magnetic field does not show any signs of decay on the timescales considered here,  either due to numerical effects or to natural relaxation of magnetic fields. While the bubble maintains its integrity, the cosmic rays gradually diffuse out of it.  However, this process is not isotropic. In the second column we see that the spatial distribution of cosmic rays initially tends to follow the distortions in the magnetic field generated by the upward motion of the bubble (lower panel, second column).  Magnetic fields and cosmic rays follow the velocity vortex pattern of the rising bubble.  At later times (upper row), the magnetic fields inside the bubble ``open up''. This,  however, takes place only in the wake of the rising bubble. The working surface of the bubble is efficiently protected from disruption by a thin layer of ordered magnetic fields (see R07 for more detailed discussion of how the field topology changes due to bubble motion). These ordered fields not only prevent instability from developing but also stop cosmic rays from escaping the bubble. On the other hand,  magnetic fields stretch out behind the bubble and become approximately parallel to the direction of the motion. Cosmic rays can therefore diffuse more easily in the direction of gravity. This effect is clearly seen in the upper row of the second column.  In the last column of Figure 1 we show what happens when diffusion is isotropic (diffusion rates in the directions perpendicular and parallel to the magnetic field lines are identical and equal to the parallel diffusion rate shown in the second column).  In this case, the diffusion time is much shorter than the buoyancy time and the pattern of energy density in cosmic rays is markedly different from that corresponding to anisotropic diffusion. In the initial stages of the evolution of CR energy density a wave-like pattern is clearly seen -- the diffusion signal propagates at finite speed.\\
\indent
Figure 2 shows the same quantities as Figure 1 but for the case where the fiducial magnitude of the diffusion coefficient $K$ is reduced by a factor of ten. The viscosity coefficient $\nu$ was increased to 0.2 to prevent numerical instabilities for the runs presented in this figure. The arrangement of panels in this figure is the same as in Figure 1. The density evolution is practically identical. The comparison between the second column with its equivalent in Figure 1 shows that, as expected, cosmic rays are now better confined to the bubbles (compare the pattern shapes and the CR intensities). The third column shows that although diffusion is formally isotropic ($K_{\perp}=K_{\parallel}$), the diffusion rate is slow enough for the fluid to advect
cosmic rays further away from the cluster centre.\\
\indent
In Figure 3 we quantify the above results in terms of the integrated cosmic ray energy normalized to the total initial energy inside the bubbles as a function of time. Dashed lines correspond to the high diffusivity case (cf. Figure 1) and solid lines to low CR diffusivity (cf. Figure 2). In each of these two cases, the lower curve corresponds to the isotropic case and the upper one to the case of anisotropic diffusion. These curves clearly demonstrate that the anisotropy effects introduced by the topology of the magnetic field are crucial in determining the escape fractions of cosmic rays from AGN bubbles.  Anisotropic diffusion can lead to a  substantial fraction of cosmic rays being retained inside the bubbles on timescales of order the buoyancy time. Any diffusion that takes place in such cases is confined to the wake of the bubble. We now move on to discuss the implications of these results for the physics of the intracluster medium. 

\newpage
\section{Astrophysical Consequences}
The above results have the following astrophysical consequences:

\begin{itemize}

\item {\it Radio relics}\\

\noindent
Our simulations show that while diffusion out of the bubbles does take place, a significant fraction of the original energy in cosmic rays can be confined to the buoyant bubbles, provided that the perpendicular diffusion coefficient is not too large. Lower values of diffusivity than those considered here would only strengthen this conclusion. If this is the case, then cosmic rays can be efficiently screened from Coulomb losses due to their interaction with the ambient medium. Such energetic plasma confined in the bubbles, when subject to passing waves, (either of AGN origin or due to infalling substructure clumps) could be re-energized and lead to detectable radio emission, thus  explaining cluster ``radio relic'' sources (En{\ss}lin \& Br{\"u}ggen 2002). The fraction of cosmic rays that does escape radio cocoons will pollute the ICM and contribute to the so called ``radio haloes'' En{\ss}lin \& Gopal-Krishna (2001).\\

\item {\it Abundance profiles in clusters of galaxies}\\

\noindent
X-ray observations (Schmidt et al. 2002, De Grandi et al. 2004, B{\"o}hringer et al. 2004; {\it in Virgo, Perseus, Centaurus} and {\it Abell 1795 clusters} reveal that clusters without cool cores show a spatially uniform distribution of metals while cooling flow clusters reveal strong centrally peaked metallicity profiles. However, the observed metallicity profiles appear much broader than the stellar light distribution of the brightest cluster galaxies (BCGs). One possible solution to this puzzle is that AGN help to uplift the gas enriched with metals (e.g., Br{\"u}ggen 2002, Roediger et al. 2007). In the absence of any microscopic energy transfer between the interior of the bubble filled with relativistic plasma and the displaced metal-enriched thermal gas, the latter will tend to sink back toward the cluster center. This problem is closely related to the problem of the development of hydrodynamical instabilities on the bubble-ICM interface. In the absence of any mechanism to suppress such instabilities (such as viscosity or magnetic fields), the rising bubble will be quickly shredded and the non-thermal bubble material will mix with the ICM, adding entropy to the metal-enriched gas and preventing it from sinking toward the cluster center. This is what happens in pure hydrodynamical simulations. In a realistic situation, magnetic fields may help to prevent such instabilities and the entropy gain of the metal-enriched gas must have a different origin. Here we suggest that the diffusion of cosmic rays out of the bubbles and their subsequent thermalization may provide the necessary energy injection.  Depending on the energy of radiating electrons and the density of the ICM, either Coulomb interaction of CR electrons with the ICM ($t_{\rm cool}^{Coul}\sim 2\times 10^{8}\gamma_{2}n_{-2}^{-1}$ yr)  or synchrotron/inverse Compton energy losses will dominate ($t_{\rm cool}^{IC}\sim 2\times 10^{8}\gamma_{4}^{-1}[1+u_{B}/u_{IC}]$ yr), where $n_{-2}$ is the electron density in units of $10^{-2}$ cm$^{-3}$,  $\gamma_{2,4}$ is the relativistic gamma factor in units of $10^{2}$ and $10^{4}$, respectively, and $u_{B,IC}$ are the energy densities in magnetic fields and CMB photons, respectively. The thermalization of CR electrons by Coulomb interactions may then preferentially increase the entropy of the metal enriched material being pulled up in the cluster atmosphere by buoyantly rising bubbles. This is because cosmic rays tend to preferentially diffuse in the direction opposite to their motion, i.e., their thermalization will mostly take place in the metal enriched ICM dragged behind the bubbles  (at least for the magnetic field parameters considered here). The observations of high abundance ridges displaced by fossil bubbles reported by Sanders, Fabian \& Dunn  (2005) are generally consistent with this idea.\\

\item {\it Excitation of H$\alpha$ line in the filaments trailing behind bubbles}\\

\noindent
Emission-line nebulae are commonly found surrounding massive galaxies in the centres of cool cluster cores (Crawford et al. 1999). The best examples of extended H$\alpha$-emitting filaments pointing from the cluster centre to the buoyant bubbles are those in the Perseus cluster (Hatch et al. 2006). Several mechanism have been proposed (and ruled out) to explain the excitation mechanism of this line. One possibility is that these nebulae are excited by stellar UV, but there is no spatial correlation between the filaments and stellar clusters (Hatch et al. 2006).  Another mechanism could be excitation by X-rays from the ICM but the filaments are up to a hundred times less luminous in X-rays than in the UV (Fabian et al. 2003b). Finally, it has been proposed that heat conduction from the ICM to the colder filaments could provide line excitation (Donahue et al. 2000). However, thermal conductivity would also lead to the evaporation of these filaments (Nipoti \& Binney 2004). Here we suggest the possibility  that cosmic rays, preferentially diffusing along the field lines trailing behind the rising bubbles, could provide the excitation mechanism for the filaments that are also located in the bubble wakes. We note that the same magnetic fields that act like ``wires'' conducting cosmic rays could also prevent the filaments from evaporating via thermal conductivity by locally suppressing it.  As a side note we add that while the ICM may be viscous, the laminar appearance of the filaments does not necessarily imply that the flow is viscous. The ordered and amplified magnetic fields trailing behind the bubble may prevent the destruction of the filaments by turbulent motions.\\

\item {\it $\gamma$-ray signatures}\\

\noindent
It has recently been proposed that extended radio features of radio galaxies (Sambruna et al. 2007; in case of Fornax A) and the interaction of cosmic rays from AGN cavities with the ICM (Hinton \& Domainko 2007; Hydra A) could be detectable with GLAST. Moreover, it has been pointed out that the interaction of such CR with the ICM could lead to significant annihilation line emission (Furlanetto \& Loeb 2002, Griffiths 2005). We point out that our simulations add more weight to this idea and that a preferential diffusion of CR in the wakes of rising bubbles could potentially lead to observable $\gamma$-ray signatures.

\end{itemize}

We note also that these results emphasize the role of non-ideal hydrodynamical effects in the studies of AGN feedback in general.  While we argued here that cosmic ray diffusivity may be one of the key effects that needs to be included, other transport process such as conduction (Ruszkowski \& Begelman 2002, Bruggen et al. 2005, Fabian et al. 2005) or viscosity (Fabian et al. 2003a, Ruszkowski et al. 2004a,b, Reynolds et al. 2005, Sijacki \& Springel 2006) may also play an important role.  Finally, if the ``pollution'' of the intracluster medium by cosmic rays diffusing out of AGN bubbles is significant, then other mechanisms such as the generalization of the magnetothermal instability (Balbus 2004, Parrish \& Stone 2005) by Chandran (2005) may also play a role.  We note that recent X-ray observations with {\it Chandra} and XMM {\it Newton} show evidence for nonthermal particle populations in the ICM (Sanders et al. 2006, Werner et al. 2007).

\section{Acknowledgements}
MR thanks Andrew Snodin for discussions.
All simulations were performed on the Columbia supercomputer
at NASA Ames center. It is MR's pleasure to thank 
the staff of NAS, and especially Johnny Chang and Art
Lazanoff, for their their highly professional help.
The Pencil Code community is thanked for making the code publicly
available. The main code website is located at
\texttt{http://www.nordita.dk/software/pencil-code/}. 
MB acknowledges support by the Deutsche Forschungsgemeinschaft
grant BR 2026/3 within the Priority Programme ``Witnesses of Cosmic History'', and MCB acknowledges partial support ubder National Science Foundation grant AST-0307502.

\begin{figure*}
\centering
\begin{minipage}[b]{\textwidth}
\centering
\includegraphics[width=0.9\textwidth, angle=0.0]{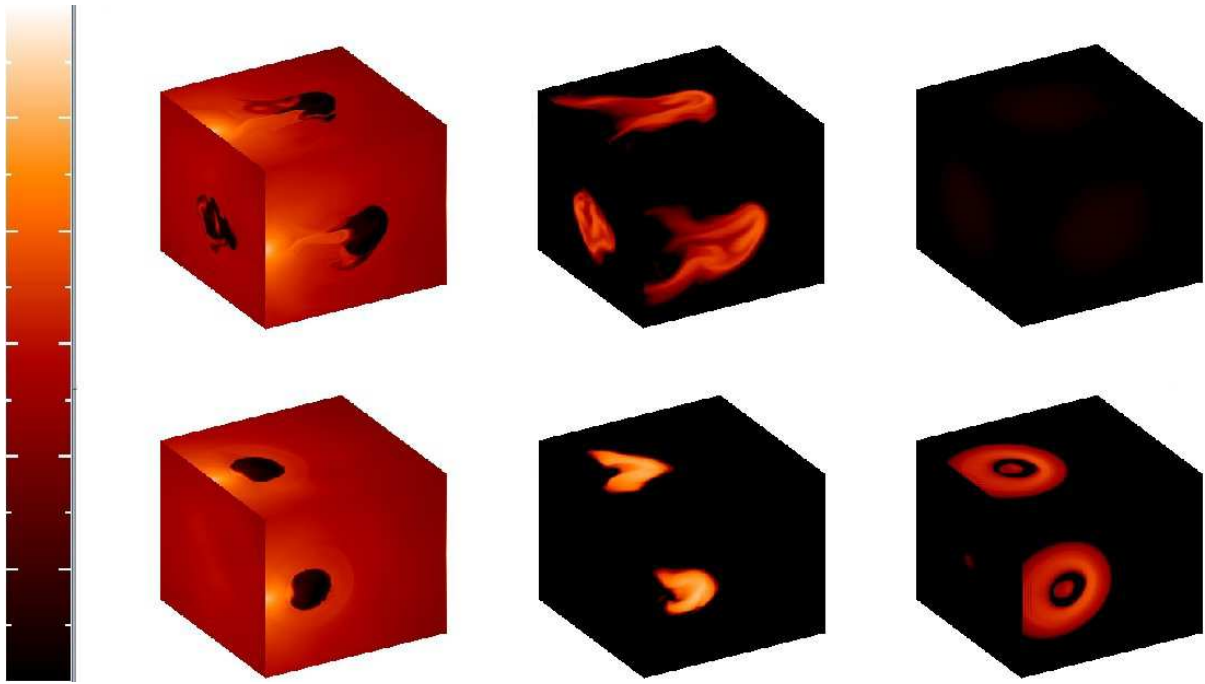}
\caption{Snapshots of density and cosmic ray energy density distributions from 3D {\it Pencil} simulations. The first column shows the natural logarithm of density, while the second and third columns present the logarithm of the energy density of cosmic rays for vanishing diffusion in the direction perpendicular to the local magnetic field ($K_{\perp}=0$, $K_{\parallel}=K$; middle column) and for the case where the diffusion of cosmic rays in the direction parallel and perpendicular to the local direction of magnetic field lines are identical ($K_{\perp}=K_{\parallel}=K$).  Lower row corresponds to earlier times $t=3$, upper to $t=25$ (one code time unit is $3.3\times 10^6$ years). In the case of density, the color bar extends from -7.0 to  -2.0 in code units. All the remaining columns are for the range [-3.0, 0.0]. See text for more details. \label{}}
\end{minipage}
\end{figure*}

\begin{figure*}
\centering
\begin{minipage}[b]{\textwidth}
\centering
\includegraphics[width=0.9\textwidth, angle=0.0]{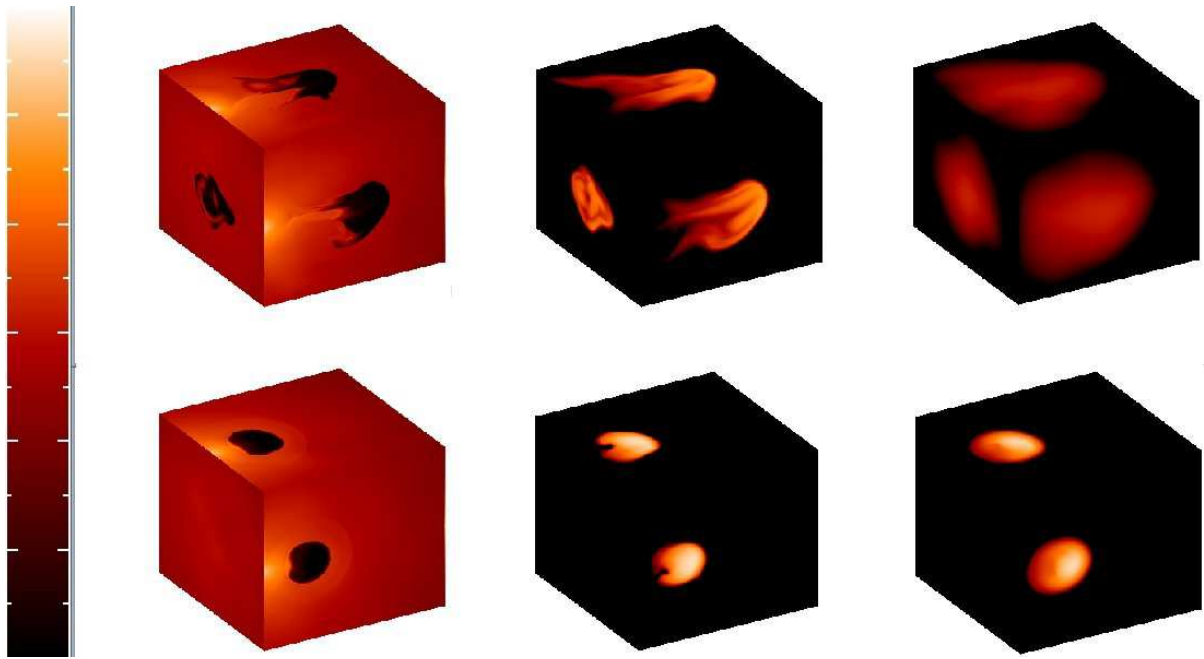}
\caption{Same as Figure 1 but for the fiducial value of diffusivity $K$ decreased by a factor of 10.\label{}}
\end{minipage}
\end{figure*}

\begin{figure}
\rotatebox{0}{\mbox{\resizebox{8.7cm}{!}{\includegraphics{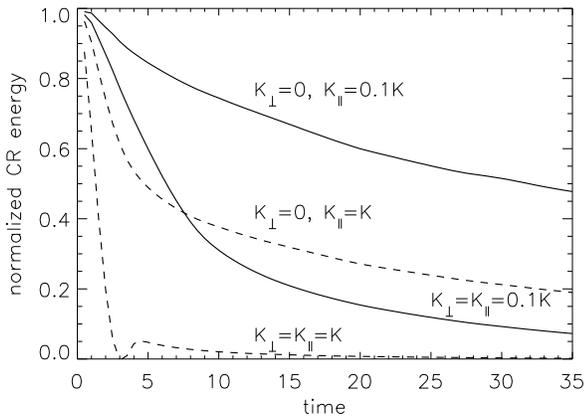}}}}
\caption{\label{}{Total energy in cosmic rays inside buoyant bubbles normalized to the total initial energy of cosmic rays as a function of time.  Dashed curves are for the high diffusivity case (cf. Figure 1) and the solid ones are for low diffusivity (cf. Figure 2). In each of these two cases, lower curves are for isotropic diffusion and the upper ones for suppressed diffusion in the direction perpendicular to the local orientation of magnetic fields. See text for more details.}}
\end{figure}

\bibliographystyle{mn2e}
\bibliography{mn}

\label{lastpage}
\end{document}